\documentclass[]{article}
\usepackage[utf8]{inputenc}
\usepackage{empheq,tikz,relsize,color,natbib}
\usepackage{graphicx,epstopdf,float,amsmath,enumitem,array,esint}
\usepackage{mathrsfs,physics,bigints,stackrel,nicematrix}

\usetikzlibrary{external}
\usepackage[margin=1in]{geometry}
\usepackage[version=4]{mhchem}
\usepackage{siunitx}
\usepackage{longtable,tabularx}
\usepackage{layouts}
\usepackage{authblk}

\newcommand{\overbar}[1]{\mkern 1.5mu\overline{\mkern-1.5mu#1\mkern-1.5mu}\mkern 1.5mu}
\newcommand{\sigmaX}[2]{\mathop{{\sum}}^{ #2}_{ #1=-#2}} 
\newcommand{\sigmaY}[2]{\mathop{{\sum}}^{ #2}_{\substack{#1=-#2,\\#1\neq 0}}} 
\makeatletter
\newcommand{\oset}[3][0ex]{\mathbin{\mathop{#3}\limits^{\vbox to#1{\kern-2\ex@\hbox{$#2$}\vss}}}}
\makeatother

\newcommand{\bfu}{{\boldsymbol{u}}} 
\newcommand{\bff}{{\boldsymbol{f}}}
\newcommand{\bkv}{{\boldsymbol{k}}}
\newcommand{\fuv}{{\boldsymbol{u}}_{\boldsymbol{k}}}
\newcommand{\ffv}{{\boldsymbol{f}}_{\boldsymbol{k}}}
\newcommand{\cHk}{{\boldsymbol{H}_{\bkv}}}
\newcommand{\cHkw}{{\boldsymbol{H}^w_{\bkv}}}
\newcommand{\cLk}{{\boldsymbol{L}_{\bkv}}}
\newcommand{\mkoL}[1]{\boldsymbol{\tilde{L}}_{#1}}
\newcommand{\mkoH}[1]{\boldsymbol{H}_{#1}}
\newcommand{\bW}[1]{{\boldsymbol{W}_{#1}}}
\newcommand{\mKI}{\boldsymbol{\boldsymbol{\tilde{I}}}}
\newcommand{\Ret}{Re_{\tau}}
\newcommand{\inv}{^{\raisebox{.2ex}{$\scriptscriptstyle-1$}}}
\newcommand{\mUU}{\boldsymbol{U}}
\newcommand{\mUUt}{U}

\newcommand{\kxx}{\kappa_x}
\newcommand{\kzz}{\kappa_z}
\newcommand{\kss}{\kappa_s}

\newcommand{\tnabla}{\tilde{\nabla}}
\newcommand{\fko}{\boldsymbol{F}}
	
\newcommand{\ucompT}[1]{{\bfu}^{s}_{#1}}
\newcommand{\fcompT}[1]{{\bff}^{s}_{#1}}
\newcommand{\wbu}{\boldsymbol{\hat{W}}}

\title{Resolvent Analysis for Turbulent Channel Flow with Riblets}

\author[1]{Andrew Chavarin\footnote{Graduate Student, Department of Aerospace and Mechanical Engineering}}
\author[2]{Mitul Luhar\footnote{Assistant Professor, Department of Aerospace and Mechanical Engineering}}
\affil[1,2]{University of Southern California, Los Angeles, CA, 90089}
\date{}
\begin{document}
\maketitle

\begin{abstract}
This paper extends the resolvent formalism for wall turbulence proposed by \citet{mckeon_sharma_2010} to account for the effect of streamwise-constant riblets. Under the resolvent formulation, the Navier-Stokes equations are interpreted as a forcing-response system: the nonlinear convective term is interpreted as a feedback forcing on the remaining linear terms, which generates a velocity and pressure response. A gain-based decomposition of the linear forcing-response transfer function --- the resolvent operator --- yields highly amplified velocity and pressure modes, which can be considered key building blocks of the turbulent flow field. Previous work has shown that these high-gain modes provide substantial insight into turbulence statistics, structure, and control of smooth-walled flows. To introduce the effect of riblets within this framework, a linear spatially-varying body force is added to the governing equations. In other words, volume penalization is used to approximate the surface features. Predictions for spanwise-periodic and streamwise-constant riblets show that specific high-gain modes identified from the modified governing equations reproduce observations made in prior direct numerical simulations with limited computation. The deterioration in performance with increasing riblet size is predicted well and so is the emergence of spanwise rollers resembling Kelvin-Helmholtz vortices. This new modeling framework is also used to pursue limited riblet shape optimization.
\vfill
\end{abstract}

\section{Nomenclature}

{\renewcommand\arraystretch{1.0}
\noindent\begin{tabular}{@{}l @{\quad=\quad} l@{}}
$x,y,z$  		    	               & Streamwise, wall-normal, spanwise directions (respectively) \\
$t$ 	 		    	               & Time \\
$\rho$ 	 		    	               & Fluid density \\
$\nu$ 	 		    	               & Kinematic viscosity \\
$\bfu = (u,v,w)$					   & Velocity vector \\
$p$									   & Pressure \\
$\mUU = (U,0,0)$   	                   & Time-averaged (mean) velocity \\
$U_n$						   		   & $n$\textsubscript{th} spatial Fourier component of mean streamwise velocity\\
$u^\prime$, $v^\prime$, $w^\prime$     & Velocity fluctuations  \\  
$P$   		    	                   & Time-averaged (mean) pressure \\
$p^\prime$     						   & Pressure fluctuations  \\  
$\kxx, \kzz$		    	           & Streamwise and spanwise wavenumber \\
$\lambda_x, \lambda_z$  	  	       & Streamwise and spanwise wavelength  \\
$\omega$							   & Frequency \\
$c = \omega/\kxx$					   & Phase speed \\ 
$\bkv = (\kappa_x,\kappa_z,\omega)$	   & Individual wavenumber-frequency combination \\
$\fuv$						           & Velocity fluctuations for Fourier mode $\bkv$\\
$\boldsymbol{W}_u$, $\boldsymbol{W}_f$ & Chebyshev weighting matrices\\
$\sigma_{\bkv,m}$					   & Rank-$m$ singular value (amplification) for Fourier mode $\bkv$\\
$\phi_{\bkv,m}$					       & Rank-$m$ right singular vector (forcing mode) at $\bkv$\\
$\psi_{\bkv,m}$					       & Rank-$m$ left singular vector (response mode) at $\bkv$\\
$u_\tau$       		    		       & Friction velocity \\
$\Ret$       		    		       & Friction Reynolds number \\
$K$		 	    	    	       	   & Dimensionless permeability \\
$s, h$								   & Riblet spacing and height \\
$A_g$								   & Cross-sectional area of riblet grooves \\
$\kss = 2\pi/s$						   & Wavenumber corresponding to riblet spacing \\
$s^+$, $h^+$					       & Inner-normalized riblet spacing and height \\
$l_g^+ = \sqrt{A_g}^+$				   & Inner-normalized groove area length scale \\
$\alpha$							   & Ridge-angle for triangular and trapezoidal riblets
\end{tabular}}

\section{Introduction}
{M}{any} turbulent flows of engineering interest involve complex patterned or rough surfaces that can substantially alter the near-wall turbulence. In most instances, such surfaces lead to an increase in turbulence and skin friction, as is the case with biofouled ship and submarine hulls \cite{schultz_bio_fouling}. However, it is well known that \textit{specific} surface features --- sharkskin-inspired riblets --- have the potential to suppress near-wall turbulence and reduce skin friction. 

Given the economic and performance advantages associated with such passive turbulence control, there have been many prior research efforts considering the effect of riblets on wall-turbulence \cite{walsh1984optimization,luchini1991resistance,choi1993direct,bechert1997experiments,gad2007flow,garcia2011drag}. Laboratory experiments performed by Walsh, Bechert, and colleagues \cite{walsh1982turbulent,walsh1984optimization,bechert1997experiments,bechert2000experiments} have created extensive databases that characterize the effect of triangular, V-shaped, scalloped, and blade-like riblets on turbulent skin friction. Drag reductions as large as $10\%$ have been measured for specific riblet geometries and sizes. Most previous studies have studied the effect of streamwise-constant riblets (i.e., riblets with constant cross-sections). However, some idealized three-dimensional geometries have also been considered \cite{bechert2000experiments,mcclure2017design}. Moving beyond laboratory experiments, the performance of V-shaped riblets has also been tested in operational conditions on aircraft with some success \cite{walsh1989riblet}.

The physical mechanism through which grooved or corrugated surfaces like riblets reduce skin friction is reasonably well understood. Riblets provide much greater resistance to the cross-flows arising from near-wall turbulence compared to the streamwise mean flow. This pushes the quasi-streamwise vortices prevalent in the near-wall region \cite[see e.g.,][]{robinson1991coherent, hamilton1995regeneration, jimenez_pinelli_1999} further away from the wall and weakens them, effectively creating a zone of limited or no turbulence within the riblet grooves \cite{chu_karniadakis_1993,choi1993direct,Sirovich1997,lee2001flow}. A weakening of the quasi-streamwise vortices hinders the turbulent transfer of mean momentum towards the wall and reduces skin friction. To quantify this effect, \citet{bechert1989viscous} suggested the use of a \textit{protrusion height}, defined as the offset between the virtual origin for the mean flow and some measure of the average wall location. Subsequent theoretical work by \citet{luchini1991resistance} showed that a better measure of the protrusion height is the difference between the virtual origin for the streamwise and spanwise flows, which can be computed via the Stokes equations for flow along and across the riblet grooves, respectively. The protrusion height computed in this fashion has proven to be a good indicator of performance for many different riblet geometries. For example, experiments conducted by \citet{bechert1997experiments} show that the protrusion height predicts the initial \textit{linear} decrease in skin friction with riblet size well for blade-shaped and scalloped riblets. However, this initial linear slope is over-predicted slightly for triangular riblets.

Although skin friction decreases linearly with increasing riblet size at first, experiments and simulations both show that performance saturates beyond a geometry-dependent size threshold and eventually degrades \cite{bechert1997experiments,garcia2011drag}. In other words, very large riblets lead to an increase in skin friction. The experimental data collected by \citet{bechert1997experiments} show that the optimal spacing for many different types of riblets ranges between $s^+ \approx 10 - 20$. More recent studies \cite{garcia2011drag,garcia2011hydrodynamic} have demonstrated that the optimal riblet size corresponds closely to $l_g^+ = \sqrt{A_g}^+ \approx 10.7$, in which $A_g$ is the cross-sectional area of the riblet grooves. Following standard notation, throughout this paper a superscript $+$ denotes normalization with respect to friction velocity, $u_\tau$, and viscosity, $\nu$. 

The observed deterioration of performance with increasing riblet size is associated with the breakdown of the viscous flow regime within the grooves or, equivalently, the onset of turbulence penetration into the grooves. Multiple different mechanisms have been proposed to explain this breakdown. Early studies suggested that the deterioration of performance may be attributed to the lodging of near-wall vortices in the riblet grooves \cite{suzuki1994turbulent,lee2001flow} or the creation of secondary streamwise vortices at the riblet tips \cite{goldstein1998secondary}. Direct numerical simulations (DNS) performed by \citet{garcia2011hydrodynamic} show that turbulent flows over riblets may also be susceptible to a Kelvin-Helmholtz type instability. More specifically, \citet{garcia2011hydrodynamic} observed the emergence of spanwise-coherent rollers with streamwise wavelength $\lambda_x^+ \approx 150$ over rectangular riblets with spacing greater than $s^+ \approx 24$ ($l_g^+\approx 15$). This threshold riblet size corresponded closely to conditions in which drag reduction performance started deteriorating. Further analysis of spectral content showed that the emergence of these spanwise-coherent structures led to a substantial increase in the Reynolds shear stress near the riblet interface. In other words, these rollers increased the turbulent transfer of momentum into the riblet grooves and thereby contributed to the breakdown of the viscous flow regime. \citet{garcia2011hydrodynamic} also carried out an inviscid linear stability analysis for the riblet flow. This analysis confirmed the emergence of a Kelvin-Helmholtz instability, though the most unstable structures were predicted to have a streamwise wavelength of $\lambda_x^+ \approx 80$. 

Despite the substantial advances made in our understanding of flows over riblets in the past three decades, there are few computationally-efficient predictive models that can be used for design and optimization of riblet geometries. The protrusion height concept provides insight into how a given riblet geometry might perform in the linear viscous regime but not at the point of maximum drag reduction. Instability analyses are able to predict the onset of performance deterioration but they cannot reproduce spectra. Moreover, to our knowledge, there are no models that can predict the degree to which riblets damp the energetically-important streaks and streamwise vortices associated with the near-wall cycle \citep{robinson1991coherent, hamilton1995regeneration, jimenez_pinelli_1999}. To address these limitations, in this paper we extend the resolvent formalism for wall-bounded turbulent flows \cite{mckeon_sharma_2010,mckeon2017engine} to account for patterned or corrugated riblet-type walls.

Low-order models based on the resolvent formalism have previously been shown to reproduce key statistical and structural features of turbulent flows over smooth walls \citep{sharma2013coherent,moarref2013model}. In recent years, resolvent analysis has also been used to develop computationally-efficient models that can be used to assess and design turbulence control techniques \citep{luhar2014opposition,luhar2015framework,luhar2016design,nakashima2017assessment}. Briefly, resolvent analysis interprets the Fourier-transformed Navier-Stokes equations (NSE) as a forcing-response system. At each wavenumber-frequency combination, a singular value decomposition of the forcing-response transfer function -- the resolvent operator -- is used to identify highly-amplified flow structures, which we term \textit{resolvent modes}. Specific high-gain resolvent modes have proven to be useful models for dynamically-important flow features such as the near-wall cycle \citep{mckeon_sharma_2010,sharma2013coherent}. As a result, these modes can also serve as low-order models for the assessment and optimization of control techniques. In other words, as a starting point, the effect of any control technique can be evaluated just on individual high-gain resolvent modes instead of the full turbulent flow field. This approach has been used successfully in the past to assess the effect of compliant walls \citep{luhar2015framework,luhar2016design} as well as active feedback flow control \citep{luhar2014opposition,nakashima2017assessment}. 

To enable direct comparison with previous DNS \citep{garcia2011drag}, the resolvent-based modeling framework developed here focuses on a fully-developed turbulent channel flow geometry. Similar models can also be developed for turbulent pipe and boundary layer flows (neglecting the slow variation in the streamwise direction). The remainder of this paper is structured as follows. Section~\ref{sec:resolvent} provides a brief review of the resolvent formulation. Section~\ref{sec:modeling_approach} describes the volume penalization technique used to account for the presences of riblets, the resulting effect on the gain-based decomposition, as well as the numerical implementation. Model predictions are presented in Section~\ref{sec:predictions}. In particular, we consider the effect of rectangular riblets of varying size on the energetic near-wall cycle (Section~\ref{sec:nw_cycle}) and we test for the emergence of spanwise-constant rollers resembling Kelvin-Helmholtz vortices (Section~\ref{sec:kh}). We also compare these predictions against DNS results \cite{garcia2011drag,garcia2011hydrodynamic}. In Section~\ref{sec:riblet_optimization}, we pursue limited optimization of rectangular, triangular, and trapezoidal riblets, and compare our predictions against prior experiments. Conclusions are presented in Section~\ref{sec:conclusion}.

\section{Resolvent Analysis}\label{sec:resolvent}
This section provides a brief review of the resolvent formulation for smooth-walled turbulent channel flow in terms of primitive variables (i.e., velocity and pressure). Further details on numerical implementation can be found in \citet{luhar2015framework}. For an in-depth discussion of the resolvent formulation for wall turbulence and related modeling efforts, the reader is referred to \citet{mckeon2017engine}.

For turbulent channel flow that is stationary in time ($t$) and homogeneous in the streamwise ($x$) and spanwise directions ($z$), the NSE can be Fourier-transformed and expressed compactly as:
\begin{equation}\label{eqn:smoothHK}
\begin{bmatrix}
\fuv \\  {p}_{\bkv}
\end{bmatrix} = \left(-i\omega \mqty[\dmat{\boldsymbol{I},0}] -\mqty[\cLk & -\tnabla \\ -\tnabla^T & 0]\right)^{-1} \mqty[\boldsymbol{I} \\ 0]\ffv = \cHk\ffv.
\end{equation}
In the expression above, $\bf\tnabla^T$ and $\bf\tnabla$ represent the divergence and gradient operators, respectively. The differential operator $\tnabla$ is defined as $\tnabla=(i\kxx,\pdv*{}{y},i\kzz)$. The first row in (\ref{eqn:smoothHK}) corresponds to the momentum equations and the second row represents the continuity constraint. A subscript $\bkv = (\kxx,\kzz,\omega)$ denotes a specific wavenumber-frequency combination. Note that $\ffv = -(\bfu\cdot\nabla\bfu)_{\bkv}$ is the Fourier-transformed nonlinear term. This nonlinear term is interpreted as a forcing that is mapped to a velocity ($\fuv$) and pressure (${p}_{\bkv}$) response by the resolvent operator, $\cHk$. Bear in mind that the resolvent depends on $\cLk$, which represents the linear terms in the NSE after Reynolds-averaging. As a result, construction of $\cHk$ requires knowledge of the turbulent mean profile, $U(y)$.

A major contribution of the resolvent analysis framework lies in the finding that the forcing-response transfer function tends to be low-rank at wavenumber-combinations that are energetic in wall-bounded turbulent flows \citep{mckeon_sharma_2010,moarref2013model}. As a result, for many $\bkv$ combinations, the discretized resolvent operator can be approximated well via a rank-1 truncation after a singular value decomposition (SVD). To ensure grid independence and enforce an $L^2$ energy norm under the SVD, the resolvent operator is scaled such that
\begin{gather}\label{eqn:resolvent_weighting}
\begin{bmatrix} \boldsymbol{W}_u& 0 \end{bmatrix}
\begin{bmatrix}
\fuv \\  {p}_{\bkv}
\end{bmatrix} = \left(\begin{bmatrix} \bW{u} & 0 \end{bmatrix}\cHk \bW{f} \inv \right) \bW{f} \ffv,
\end{gather}
or
\begin{equation}
\bW{u}\,\fuv = \cHkw\,\bW{f} \ffv.
\end{equation}
The block diagonal matrices $\bW{u}$ and $\bW{f}$ contain numerical quadrature weights which ensure that the SVD (and rank-1 truncation) of the scaled resolvent operator 
\begin{equation}
\cHkw= \sum_{m} \psi_{\bkv,m} \sigma_{\bkv,m} \phi^{\ast}_{\bkv,m} \approx \psi_{\bkv,1}\sigma_{\bkv,1} \phi^{\ast}_{\bkv,1}
\end{equation}
where
\begin{gather}
\sigma_{\bkv,1} > \sigma_{\bkv,2}...  > \sigma_{\bkv,m}... >0, \:\:
{\phi^*_{\bkv,l}}{\phi^{}_{\bkv,m}} = \delta_{lm}, \:\: 
{\psi^*_{\bkv,l}}{\psi^{}_{\bkv,m}} = \delta_{lm},
\label{eqn:SVD}
\end{gather}
yields forcing modes ${\boldsymbol{f}_{\bkv,m}} =\bW{f}^{-1} {\phi_{\bkv,m}}$ and velocity response modes ${\bfu_{\bkv,m}} = \bW{u}^{-1} {\psi_{\bkv,m}}$ with unit energy over the channel cross-section. In other words, the scaling ensures that the orthonormality constraints in (\ref{eqn:SVD}) yield
\begin{equation}\label{eqn:SVDNorm}
\int_{0}^{2} \boldsymbol{f}^*_{\bkv,l} \boldsymbol{f}^{}_{\bkv,m} \,dy = \delta_{lm} \:\: , \:\: \int_{0}^{2} \bfu^*_{\bkv,l} \bfu^{}_{\bkv,m} \,dy = \delta_{lm},
\end{equation}
where a superscript $*$ denotes a complex conjugate. The wall-normal $y$ coordinate is normalized with the channel half-height, $h$; $y=0$ denotes the lower wall of the channel and $y=2$ represents the upper wall. 

Equations (\ref{eqn:smoothHK})-(\ref{eqn:SVD}) show that forcing in the direction of the $m^{th}$ singular forcing mode with unit amplitude results in a response in the direction of the $m^{th}$ singular response mode that is amplified by the singular value $\sigma_{\bkv,m}$. Thus, forcing $\boldsymbol{f}_{\bkv}=\boldsymbol{f}_{\bkv,1}$ creates a response $\bfu_{\bkv} = \sigma_{\bkv,1} \bfu_{\bkv,1}$. As noted earlier, the resolvent operator is known to be low-rank at $\bkv$ combinations that are energetic in natural turbulence, often with $\sigma_{\bkv,1} \gg \sigma_{\bkv,2}$. High amplification of the rank-1 response modes implies that the flow field may be reasonably approximated using $\bfu_{\bkv,1}$ at each $\bkv$. Recent studies show that this rank-1 approximation is able to reproduce key structural and statistical features of smooth-walled turbulent flows \citep{moarref2013model,mckeon2017engine}. This rank-1 truncation has also yielded useful control-oriented low-order models \citep{luhar2014opposition,luhar2015framework,luhar2016design,nakashima2017assessment}. In particular, it has been demonstrated that the amplification of specific rank-1 response modes that resemble dynamically-important features of wall-bounded turbulent flows (e.g., the near-wall cycle) is a useful predictor of performance. Motivated by these previous modeling efforts, we retain the rank-1 approximation for the remainder of this paper. We assume that the flow field at each $\bkv$ corresponds to the rank-1 response $\bfu_{\bkv,1}$. The term \textit{resolvent modes} refers to these rank-1 flow fields. The terms amplification and gain are used interchangeably to refer to the corresponding singular values, $\sigma_{\bkv,1}$.

\section{Modeling Approach}\label{sec:modeling_approach}

\subsection{Volume Penalization for Riblets}\label{sec:wall_model}
To account for the effect of patterned walls of various geometries and shapes we employ a volume penalization technique: the \textit{fictitious domain} methodology of \citet{khadra2000}. Solid obstructions within the fluid domain --- riblets in our case --- are modeled as a spatially varying permeability function $K(x,y,z)$, which appears as an additional body force in the momentum equations. This approach allows consolidation of the coupled fluid-solid system into a single heterogeneous numerical domain. The equations governing fluid flow in this domain are given by:
\begin{gather}
\pdv{\bfu}{t} + \bfu \cdot \nabla \bfu = -\nabla p + \frac{1}{\Ret}\nabla^2 \bfu - K\inv \bfu, \nonumber \\
\nabla \cdot \bfu = 0.
 \label{eqn:governing_VANS} 
\end{gather}
The volume penalization term is modeled using a Darcy-type linear resistance, with dimensionless permeability $K$. The friction Reynolds number is defined as $\Ret = u_\tau h/\nu$, in which $u_\tau$ is the friction velocity and $\nu$ is kinematic velocity. 

For our problem, the permeability function takes on two values. In the fluid region that is free from all solid obstacles, $K\rightarrow \infty$. Here, the governing equations reduce down to the standard NSE. In the region of the domain that is occupied by a solid obstruction, $K \rightarrow 0$. Here, the permeability term becomes the lead order term in Eq.~\eqref{eqn:governing_VANS}, forcing the velocity in the solid domain towards zero. \citet{angot1999analysis} provided a rigorous analysis of the error bounds for this method. The $H^1$ norm of the error in velocity computed by the penalization method, when compared to the true velocity field, is expected to be $\mathscr{O}(K^{1/4})$ over the entire domain. The $L^2$ norm of error in velocity is expected to be $\mathscr{O}(K^{3/4})$ in the solid domain and $\mathscr{O}(K^{1/4})$ in the fluid domain \cite{angot1999analysis,khadra2000}. Since $K$ cannot be driven to zero numerically, a value of $K \sim \mathscr{O}(10^{-8})$ guarantees an $L^2$ error of $\mathscr{O}(10^{-2})$ in the fluid region of the domain, and of $\mathscr{O}(10^{-5})$ in the solid region. For completeness, we note that the volume penalization approach is conceptually similar to more sophisticated immersed boundary techniques used in fluid flow simulations, albeit without a feedback component to the body force that explicitly enforces the boundary conditions \citep{goldstein_handler_sirovich_1995,goldstein1998secondary,peskin2002immersed}.

This volume penalization method allows us to avoid the additional work required to form meshes that conform to the geometry of the riblets. Instead, we utilize a simpler Cartesian mesh corresponding to the standard channel flow geometry into which the riblets are embedded via the modified governing equations. From the perspective of resolvent analysis, the key benefit of this method is that the modified equations retain the linearity of the forcing-response transfer function used for the gain-based decomposition. In other words, the resolvent analysis approach described in the previous section can be extended to account for the effects of riblets as long as the permeability function can be expressed analytically and its spatial Fourier transform can be computed.

\subsection{Modified Resolvent Operator}\label{sec:modified_resolvent}
For the remainder of this paper we focus on streamwise-constant riblets such that the permeability function only varies in the wall-normal and spanwise directions, $K(y,z)$. Moreover, assuming that the riblets are periodic in the spanwise direction with spacing $s$, the permeability function can be expressed analytically as a Fourier series of the form
\begin{equation}\label{eqn:wall_model}
K\inv(y,z) = \sigmaX{n}{\infty}a_n(y)\exp{in\kss z}
\end{equation}
with $\kss = 2\pi/s$. Thus, the riblets are modeled as a series of permeability \textit{pulses} stacked in the wall-normal direction. The spacing between the pulses is always $s$. The wall-normal variation in the width of these pulses is set by the Fourier coefficients $a_n(y)$. Note that pulse width determines the riblet geometry. As an example, for rectangular riblets, the pulse width is constant in $y$ over the height of the riblets. For triangular or trapezoidal riblets, the pulse width decreases linearly over the height of the riblets. Since the mean velocity profile will exhibit the same periodicity as the riblets, this profile can also be expressed as a Fourier series of the form:
\begin{equation}\label{eqn:mean22_fourier}
U(y,z) = \sigmaX{n}{\infty} U_n(y) \exp{in\kss z}. 
\end{equation}

To construct the resolvent operator, we follow a similar set of steps to those outlined in Section~\ref{sec:modeling_approach}. The full turbulent velocity field in the channel is decomposed into a time-averaged mean component and fluctuations about this mean
\begin{equation}
\bfu(\boldsymbol{x},t) = \mUU(\boldsymbol{x}) + {\bfu}^\prime(\boldsymbol{x},t).
\end{equation}
This yields the following equation for the non-zero streamwise component of the mean flow
\begin{equation}\label{eqn:mean_vel1}
- \pdv{P}{x} + \frac{1}{\Ret}\nabla^2 U - K\inv U - \left(\nabla\cdot(\overbar{\bfu^\prime{\bfu^\prime}})\right)_x = 0.
\end{equation}
The subscript $x$ for the Reynolds stress term denotes the streamwise component. The fluctuations satisfy
\begin{equation}\label{eqn:fluctuations}
\pdv{\bfu^\prime}{t} + \mUU\cdot\nabla \bfu^\prime + \bfu^\prime\cdot\nabla\mUU 
+ \nabla\cdot({\bfu^\prime}{\bfu^\prime} -\overbar{{\bfu^\prime}{\bfu^\prime}}) 
=-\nabla p^\prime + \frac{1}{\Ret}\nabla^2{\bfu^\prime} - K\inv\bfu^\prime,
\end{equation}
and the continuity constraint
\begin{equation}
\nabla \cdot \bfu^\prime =0.
\end{equation} 

To estimate the mean profile, which is required for the resolvent operator, we model the Reynolds stress term in (\ref{eqn:mean_vel1}) using an eddy viscosity, $\nu_e$. Combining (\ref{eqn:wall_model}), (\ref{eqn:mean22_fourier}), and (\ref{eqn:mean_vel1}) with the eddy viscosity model, the governing equation for the $n^{th}$ Fourier mode for the mean flow can be expressed as
\begin{equation}
\frac{1}{\Ret} \left( \left(\nu_e\dv[2]{}{y} + \dv{\nu_e}{y}\dv{}{y}\right)\delta(n) + \dv[2]{}{y} -n^2\kss^2 \right) {\mUUt}_{n} -\sum_{p+q=n}{\mUUt}_{p}a_q=\pdv{P}{x}\delta(n),\label{eqn:mean_vel2}
\end{equation}
in which $\delta(n)=1$ for $n=0$ and $\delta(n)=0$ for $n \neq 0$. As shown in (\ref{eqn:eddy_viscosity}), the eddy viscosity is assumed to act only on the spanwise-constant component of the mean flow ($U_0$). The mean pressure gradient driving the channel flow is also assumed to be spanwise-constant. Further details on the eddy viscosity formulation and the numerical procedure used to compute the coupled set of equations for each mean flow harmonic are provided in Section~\ref{sec:implementation}.

The fluctuating component of the flow field is Fourier-transformed in the homogeneous spatial directions ($x,z$) and time ($t$). Under the Fourier transform, the governing equation for a specific wavenumber-frequency combination $\bkv$ can be expressed as
\begin{gather}
\left(-i\omega + i\kxx{\mUUt}_{0} + a_0 - \frac{1}{\Ret}\tnabla^2\right)\fuv +  
\tnabla p_{\bkv} + v_{\bkv}\pdv{\mUUt_{0}}{y}\boldsymbol{e_x} \nonumber \\
+ \sigmaY{n}{\infty} \left(a_n + i\kxx{\mUUt}_{n} \right) \bfu_{\bkv-n\kss} 
+ \sigmaY{n}{\infty} \left(v_{\bkv-n\kss}\pdv{}{y}+ in\kss w_{\bkv-n\kss} \right) \mUUt_{n}\boldsymbol{e_x}
= \ffv, \label{eqn:almost_there_resolvent}
\end{gather}  
in which the subscript $\bkv-n\kss$ represents the wavenumber frequency combination $(\kxx,\kzz-n\kss,\omega)$. The top row in (\ref{eqn:almost_there_resolvent}) represents linear interactions between the Fourier mode of interest and the $0^{th}$ spatial harmonic of the mean flow and the penalization term (i.e., the spanwise-constant components). The second row includes interactions with the remaining spatial harmonics of the mean flow and penalization term. Equation (\ref{eqn:almost_there_resolvent}) can be organized into the following block operator form, grouping together the terms involving interactions with the spanwise-constant and spanwise-varying components of the mean flow and penalization term
\begin{equation}
\left(\,-i\omega 
\begin{bmatrix}
\boldsymbol{I} & \\
& {0}
\end{bmatrix}-
\begin{bmatrix}
\cLk - a_0\boldsymbol{I} & - \tnabla \\
\tnabla^T & 0
\end{bmatrix}\,\right)
\begin{bmatrix}
\fuv \\
p_{\bkv}
\end{bmatrix} 
+ \sigmaY{n}{\infty} \fko_n
\begin{bmatrix}
\bfu_{\bkv-n\kss} \\
p_{\bkv-n\kss}
\end{bmatrix} 
=
\begin{bmatrix}
\boldsymbol{I} \\0
\end{bmatrix}\ffv,\label{eqn:modified_resolvent1}
\end{equation}
in which the linear operator $\cLk$ is identical to that in (\ref{eqn:smoothHK})
\begin{equation}
\cLk =
\begin{bmatrix}
-i\kxx{\mUUt}_{0} +\Ret\inv\tnabla^2 & -{d\mUUt_0}/{dy} & 0                \\    
0	         & -i\kxx{\mUUt}_{0} +\Ret\inv\tnabla^2 &       0           \\
0            &         0         & -ik_x{\mUUt}_{0} +\Ret\inv\tnabla^2 
\end{bmatrix}, \label{eqn:modified_resolvent_Lk}
\end{equation}
and the block operator $\fko_n$ is defined as
\begin{equation}
\fko_{n} =
\begin{bmatrix}
a_n+i\kxx{\mUUt}_{n} & {d\mUUt_n}/{dy} & in \kss {\mUUt}_{n}               & 0 \\    
&a_n+ik_x{\mUUt}_{n}   & 0               & 0 \\
&                  & a_n+ik_x{\mUUt}_{n} & 0 \\
0            &                  &                 & 0  
\end{bmatrix}. \label{eqn:modified_resolvent_Fn}
\end{equation}
Equation (\ref{eqn:modified_resolvent1}) demonstrates that the spanwise variation in the governing equations and mean velocity profile introduced by the riblets couples together the wavenumber-frequency combination of interest, $\bkv$, and an infinite set of harmonics with wavenumber-frequency combinations $\bkv-n\kss=(\kxx,\kzz-n\kss,\omega)$ in which $n = \{\pm 1, \pm 2, ... \pm \infty\}$. If the wall is homogeneous, containing no spatial variation in the spanwise direction, the coupling terms drop out and resolvent analysis can proceed on a mode-by-mode basis, as shown in (\ref{eqn:smoothHK}) for the smooth-wall case. However, for the case with riblets, the coupled system must be analyzed together. 

To proceed, we truncate the coupled input-output system to only consider harmonics with $n = \{\pm 1, \pm 2, ... \pm n_h\}$. The governing equations for this system of interacting modes can be expressed as
\begin{equation}
\begin{bNiceMatrix}[nullify-dots]
\ucompT{\bkv+n_h\kss} \\ \Vdots \\ \ucompT{\bkv} \\ \Vdots \\ \ucompT{\bkv-n_h\kss}
\end{bNiceMatrix} =
\left(-i\omega
\begin{bNiceMatrix}[nullify-dots]
\mKI &        &     &       &       \\
     &\Ddots  &     &       &       \\
     &        &\mKI &       &       \\
     &        &     &\Ddots &  	    \\
     &        &     &       & \mKI
\end{bNiceMatrix}
 +
\begin{bNiceMatrix}[nullify-dots]  
\mkoL{\bkv+n\kss} & \fko_{1}  &  \Cdots      & \Cdots     & \fko_{2n_h}   \\ 
    \fko_{-1}     & \Ddots    &  		     &            & \Vdots    \\ 
    \Vdots        &           &  \mkoL{\bkv} &            & \Vdots    \\ 
    \Vdots        &           & 		     & \Ddots 	  & \fko_{1}  \\ 
    \fko_{-2n_h}  & \Cdots    &  \Cdots      & \fko_{-1}  & \mkoL{\bkv-n\kss} 
\end{bNiceMatrix}
\right)^{-1}
\begin{bNiceMatrix}[nullify-dots]
\fcompT{\bkv+n_h\kss} \\ \Vdots \\ \fcompT{\bkv} \\ \Vdots \\ \fcompT{\bkv-n_h\kss}
\end{bNiceMatrix}. \label{eqn:modified_resolvent_f}
\end{equation}
\begin{equation}
\end{equation}
Here, the superscript $s$ is used to denote the complete system of governing equations, i.e., the momentum equations as well as the continuity constraint. In other words, $\ucompT{\bkv} = [\fuv, p_\bkv]^T$ includes both the velocity and pressure fields, while the forcing term $\fcompT{\bkv} = [\ffv,0]^T$ explicitly accounts for the continuity constraint. The operators $\mKI$ and $\mkoL{\bkv}$ are defined as
\begin{equation}
\mKI =\begin{bmatrix}
\boldsymbol{I} & \\
& {0}
\end{bmatrix}\quad \text{and} \quad \mkoL{\bkv}
=-\begin{bmatrix}
\cLk - a_0\boldsymbol{I} & -\tnabla \\
\tnabla^T & 0
\end{bmatrix},
\end{equation}
respectively. Equation (\ref{eqn:modified_resolvent_f}) can be expressed more compactly as
\begin{equation}\label{eqn:modified_resolvent_compact}
\bfu^s_{\bkv+} = \mkoH{\bkv+} \bff^s_{\bkv+}
\end{equation}
with
\begin{equation}
\bfu^s_{\bkv+} =
\begin{bmatrix}
\ucompT{\bkv+n_h\kss}, & \cdots & \ucompT{\bkv}, & \cdots & \ucompT{\bkv-n_h\kss} 
\end{bmatrix}^T
\quad \text{and} \quad 
\bff^s_{\bkv+}=
\begin{bmatrix}
\fcompT{\bkv+n_h\kss}, & \cdots & \fcompT{\bkv}, & \cdots & \fcompT{\bkv-n_h\kss}
\end{bmatrix}^T
\end{equation}
The notation $\bkv+$ represents the coupled system of Fourier modes associated with the individual wavenumber-frequency combination $\bkv$. Similar to the steps outlined in (\ref{eqn:resolvent_weighting})-(\ref{eqn:SVDNorm}) for the smooth-wall case, the extended resolvent operator in (\ref{eqn:modified_resolvent_compact}) is weighted using the extended quadrature matrices, $\wbu_u$ and $\wbu_f$,
\begin{equation}
\wbu_u\bfu^s_{\bkv+} = \left(\wbu_u \mkoH{\bkv+} \wbu_f\inv\right)\wbu_f \bff^s_{\bkv+},
\end{equation}
prior to carrying out the SVD and rank-1 truncation
\begin{equation}\label{eqn:modified_resolvent_svd}
\left(\wbu_u \mkoH{\bkv+} \wbu_f\inv \right)= \sum_{m} \psi_{\bkv+,m} \sigma_{\bkv+,m} \phi^{\ast}_{\bkv+,m} \approx \psi_{\bkv+,1}\sigma_{\bkv+,1} \phi^{\ast}_{\bkv+,m}.
\end{equation}
Recall that this weighting step ensures that the forcing modes (right singular vectors, $\phi_{\bkv+,m}$) and response modes (left singular vectors, $\psi_{\bkv+,m}$) identified from the SVD have unit energy when integrated over the channel. 

The highest-gain flow structure, which comprises a coupled set of traveling wave Fourier harmonics, is computed from the rank-1 response mode via the inverse-weighting operation
\begin{equation}\label{eqn:flow_structure}
\bfu^s_{\bkv+,1} = \wbu^{-1}_u \psi_{\bkv+,1}.
\end{equation}
The singular value $\sigma_{\bkv+,1}$ is used as a measure of energy amplification for this high-gain flow structure. In Section~\ref{sec:predictions}, we show that changes in the gain and structure of specific resolvent modes provide substantial insight into riblet performance. Further details on numerical implementation (i.e., discretization in $y$ direction, grid resolution, number of coupled Fourier harmonics retained) are provided in the following section.

\subsection{Numerical Implementation}\label{sec:implementation}
The mean flow equations and resolvent operator are discretized in the wall-normal $y$ direction using the split-domain approach described in \citet{rectangular_block_operators_trefethen}. The upper domain corresponds to the unobstructed flow and is bounded between the riblet tips ($y=0$) and the channel centerline ($y=1$). The bottom domain corresponds to the region in which the riblets are present and the governing equations include the penalization term; this domain is bounded by the channel flow ($y=-h$) and the riblet tips ($y=0$). This split-domain representation ensures that the riblet height, $h$, remains fixed even as grid resolution changes. In other words, it allows us to impose a hard wall-normal cut-off for the penalization term. Across the domains, we impose the following matching conditions for each spatial harmonic of the mean velocity profile
\begin{equation}
\left.{\mUUt}_{n}\right\lvert_{y=0^+}-\left.{\mUUt}_{n}\right\lvert_{y=0^-}=0
\quad \text{and}\quad
\left.\frac{d\mUUt_n}{dy}\right\lvert_{y=0^+}-\left.\frac{d\mUUt_n}{dy}\right\lvert_{y=0^-}=0.
\end{equation}  
Similarly, the following matching conditions are used for each harmonic present in the extended resolvent operator (\ref{eqn:modified_resolvent_f})
\begin{gather}
\left.{\bfu}_{\bkv}\right\lvert_{y=0^+} -\left.{\bfu}_{\bkv}\right\lvert_{y=0^-} =0,\\
\left.{p}_{\bkv}\right\lvert_{y=0^+} -\left.{p}_{\bkv}\right\lvert_{y=0^-}=0, \\
\left(\left.\frac{du_{\bkv}}{dy} \right\lvert_{y=0^+} -\left.\frac{du_{\bkv}}{dy} \right\lvert_{y=0^-}\right)\boldsymbol{e}_x +
\left(\left.\frac{dw_{\bkv}}{dy} \right\lvert_{y=0^+} -\left.\frac{dw_{\bkv}}{dy} \right\lvert_{y=0^-}\right)\boldsymbol{e}_z=0.
\end{gather}
At the lower wall of the channel, $y=-h$, the standard no-slip boundary conditions are applied for the mean flow as well as the Fourier-transformed fluctuations.

As noted earlier, an eddy viscosity model is used to compute the mean velocity profile needed for the resolvent operator. To solve for the coupled system of mean flow harmonics governed by (\ref{eqn:mean_vel2}), we assume that the eddy viscosity is zero in the lower domain occupied by the riblets. In other words, flow within the riblet grooves is assumed to remain laminar and so the velocity profile can be computed using the penalized version of the Stokes equations. In the upper unobstructed domain, we employ the eddy viscosity model proposed by \citet{reynolds_tiederman_1967}. Thus, we have
\begin{equation}
\nu_e =\left\lbrace
\begin{matrix}\frac{1}{2}\left\lbrack 1+\left(\frac{\Ret\kappa}{3}(2y-y^2)(3-4y+2y^2)\left(1-\exp((|y-1|-1)\frac{\Ret}{\alpha})\right)\right)^2
\right\rbrack^{1/2}- \frac{1}{2} & y \ge 0\\
0& y < 0
\end{matrix}\right..\label{eqn:eddy_viscosity}
\end{equation}
The eddy viscosity shown above is normalized by the kinematic viscosity, $\nu$. The parameter $\alpha$ appears in van Driest's mixing length model and $\kappa$ is the von Karman constant.

For the results shown below, a total of $N=81$ Chebyshev nodes is used for discretization in the wall-normal direction; $2N/3 = 54$ nodes are used in the upper unobstructed domain ($y \in [0,1]$) and $N/3 = 27$ nodes are used in the lower domain occupied by the riblets ($y \in [-h,0]$). The number of Fourier harmonics is truncated to $n = \pm n_h = \pm 6$. For the mean flow, this choice of $N$ and $n_h$ led to convergence within $\mathscr{O}(10^{-2})$ in the predicted slip velocity at the riblet tips, the centerline velocity, and the bulk-averaged velocity compared to predictions made with $N = 120$ and $n_h = 10$. Singular values for the resolvent modes considered in this paper also converged to within $\mathscr{O}(10^{-2})$. Convergence to $\mathscr{O}(10^{-2})$ was deemed sufficient for the purposes of this study for two reasons. First, the $L^2$ norm of the error associated with the volume penalization technique is expected to be of $\mathscr{O}(10^{-2})$. Second, compared to the smooth-wall case, riblets led to changes in singular values that were significantly greater than $\mathscr{O}(10^{-2})$. Note that the extended resolvent operator in (\ref{eqn:modified_resolvent_f}) is of size $4N(2n_h+1)\times 4N(2n_h+1)$. As a result, increases in $N$ and $n_h$ carry a substantial computational penalty. For instance, the computation time required to compute the resolvent operator and SVD for a single set of coupled Fourier modes increases from $\approx 60$ seconds (on a single core of a desktop workstation) for $N=81$ and $n_h = 6$ to $\approx 600$ seconds for $N=100$ and $n_h = 10$.

\begin{figure}
	\centering
	\includegraphics[scale= 1]{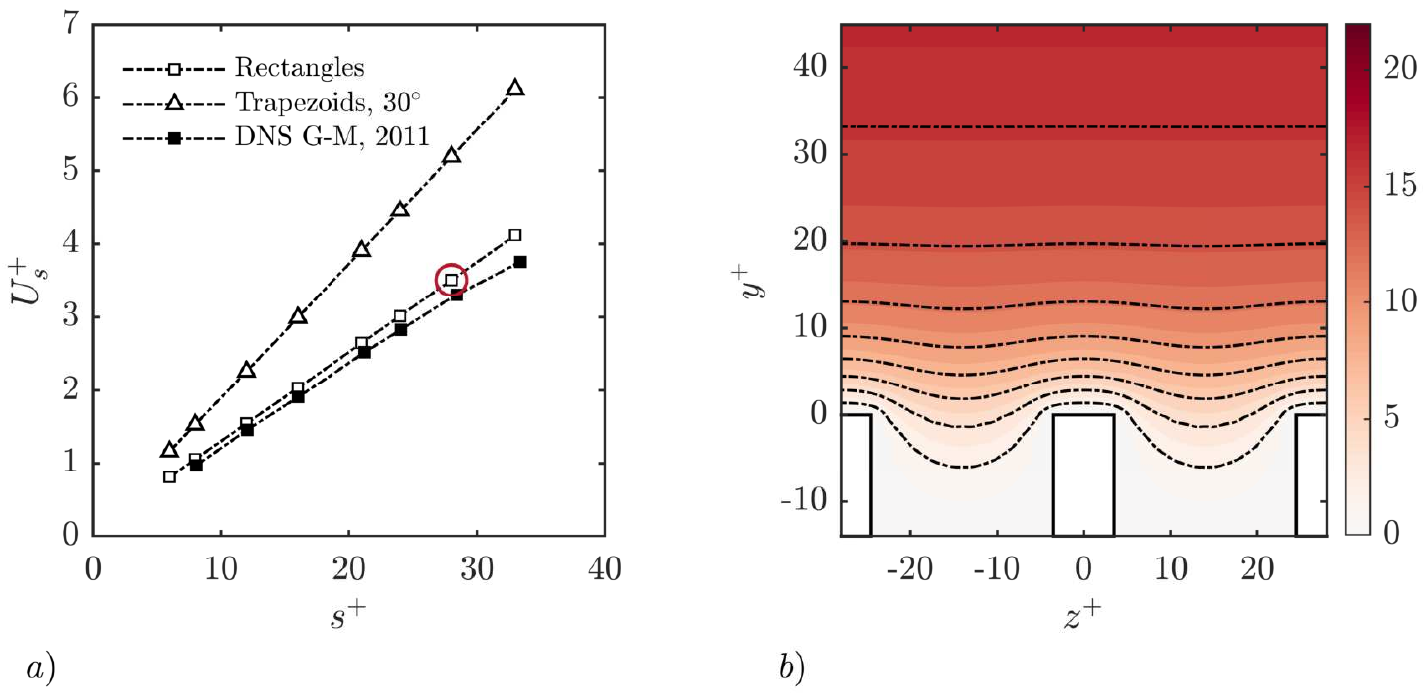}
	\caption{Mean velocity predictions. (a) Predicted slip velocity as a function of riblet spacing. Open symbols show predictions for rectangular and trapezoidal riblets. Closed symbols show DNS results from \citet{garcia2011hydrodynamic} for rectangular riblets. (b) Mean streamwise velocity profile for the case identified in (a) in red. Solid contours show velocity values in increments of $2 u_\tau$.}\label{fig:mean_velocity_profile}
\end{figure}

Sample predictions for the mean velocity profile are shown in fig.~\ref{fig:mean_velocity_profile} for rectangular and trapezoidal riblets of varying size at $\Ret = 180$. For the rectangular riblets, the riblet height is fixed at $h/s=0.5$ and width is fixed at $t/s=0.25$, in which $s$ is the spacing of the riblets. This geometry is identical to that considered by \citet{garcia2011drag} in DNS. For the trapezoidal riblets, the ridge angle is set at $\alpha = 30^\circ$. Despite the substantial simplifications (penalization, truncation to $\pm n_h$ harmonics, eddy viscosity model) the predicted slip velocity at the tips of the rectangular riblets is in good agreement with DNS results from \citet{garcia2011hydrodynamic}; see fig.~\ref{fig:mean_velocity_profile}a. Over all the configurations tested, the error in our predictions was less than $8\%$ compared to the DNS results from \citet{garcia2011hydrodynamic}. The relative error in predicted slip velocity increased with increasing riblet size. This increase in error could be due to the breakdown of the laminar flow assumption in the region occupied by the riblets. As riblet size increases, the effect of turbulence penetration into the grooves must be considered \citep{garcia2011hydrodynamic}. Further, the predicted slip velocity for trapezoidal riblets is larger than that for rectangular riblets of similar size, which is consistent with physical intuition. The sample mean profile predictions shown in fig.~\ref{fig:mean_velocity_profile}b confirm that volume penalization is able to successfully enforce the presence of riblets.

\section{Model Predictions and Discussion}\label{sec:predictions}
The literature review presented in the introduction suggests that, as a starting point, riblet performance may be assessed by (i) considering the effect of riblets on the energetic NW cycle, and (ii) by testing for the emergence of spanwise-constant rollers resembling Kelvin-Helmholtz vortices. With this in mind, we employ the extended resolvent framework to consider the effect of riblets on resolvent modes resembling the NW cycle and modes that are spanwise-constant. To enable direct comparison with the DNS results of \citet{garcia2011hydrodynamic}, we focus on geometrically-similar rectangular riblets with height $h/s = 0.5$ and thickness $t/s = 0.25$ at friction Reynolds number $\Ret = 180$. \citet{garcia2011hydrodynamic} considered riblets with inner-normalized spacing between $s^+ = 8$ and $s^+ = 33$. Here, we consider riblet spacing ranging from $s^+ = 0.5$ to $s^+=40$.

Throughout this section, we use the term gain and amplification interchangeably to refer to the first singular value: $\sigma_{\bkv+,1}$ in (\ref{eqn:modified_resolvent_svd}). Predictions for flow structure in physical space are obtained via an inverse Fourier transform of the highest-gain response modes: $\bfu^s_{\bkv+,1}$ in (\ref{eqn:flow_structure}). Note that $\bfu^s_{\bkv+,1}$ represents a coupled set of Fourier harmonics. From here on, we drop the additional subscript $1$ for convenience.

\subsection{Near-Wall Cycle}\label{sec:nw_cycle}
In smooth wall flows, the dynamically-important NW cycle is known to comprise alternating streaks of high and low streamwise velocity, with streamwise length $\lambda^+_x \approx 10^3$ and spanwise length $\lambda^+_x \approx 10^2$. In addition, these structures are found in the buffer region of the flow, near $y^+\approx 15$ where the local mean velocity is $U^+ \approx 10$ \citep[see e.g.,][]{robinson1991coherent,jimenez_pinelli_1999}. To evaluate the effect of riblets on such structures at $\Ret = 180$, we consider resolvent modes corresponding to $\bkv=(\kxx,\kzz,c^+) = (2\pi\Ret/10^3, 2\pi\Ret/10^2,10)$. Previous studies have demonstrated that resolvent modes with these length and velocity scales reproduce known features of the NW cycle \citep{mckeon_sharma_2010,sharma2013coherent}. Recent studies have also shown that these structures provide a useful starting point for the evaluation and design of control techniques \citep{luhar2014opposition,luhar2015framework,nakashima2017assessment}.

\begin{figure}
	\centering
	\includegraphics[scale = 1]{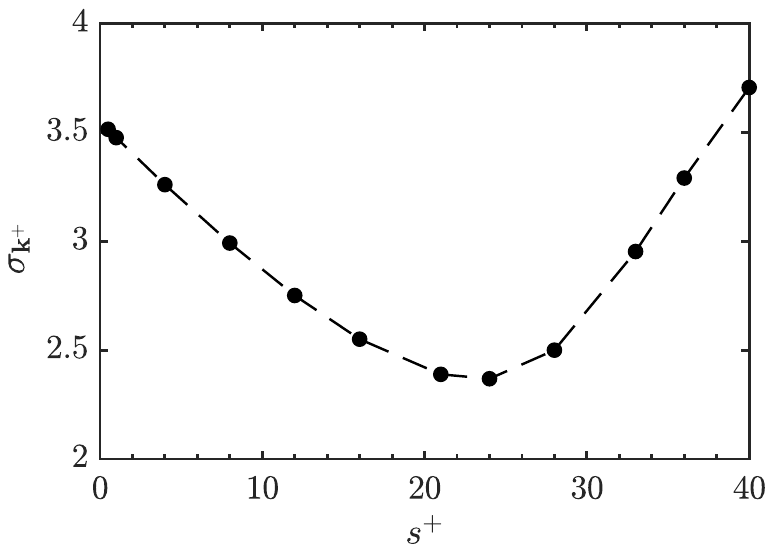}
	\caption{Gain for resolvent modes resembling the NW cycle as a function of riblet spacing.}\label{fig:nwc_for_amp}	
\end{figure}

Figure~\ref{fig:nwc_for_amp} shows how riblets of varying size affect the gain (singular value, $\sigma_{\bkv+}$) for the NW resolvent modes. For smooth wall flows at this Reynolds number, a gain-based decomposition of the unmodified resolvent operator (\ref{eqn:smoothHK}) yields a singular value of $\sigma_{\bkv} \approx 3.56$. The gain predicted by the extended resolvent operator (\ref{eqn:modified_resolvent_f}) approaches this value for riblet spacing $s^+ = 0.5$. Keep in mind that the extended resolvent operator utilizes a different wall-normal discretization compared to the smooth wall case. Further, the volume-penalized governing equations are used in the region occupied by the riblets. Given these differences, asymptotic convergence towards the smooth-wall value as $s^+ \rightarrow 0$ serves as additional verification of our modeling approach.

The gain distribution shown in fig.~\ref{fig:nwc_for_amp} also reproduces two important trends that are consistent with drag reduction curves from previous experiments and DNS. First, $\sigma_{\bkv+}$ initially decreases linearly with increasing riblet size. Since the singular values are a measure of energy amplification within the resolvent framework, this reduction in $\sigma_{\bkv+}$ can be interpreted as mode suppression due to the presence of riblets. Thus, the monotonic decrease in gain is consistent with previous observations which show that drag decreases linearly with increasing riblet size in the viscous regime \citep{bechert1997experiments,garcia2011hydrodynamic}. Second, the singular values approach a minimum value for spacing $s^+ \approx 23$, and above this value there is an increase in gain. This critical value for riblet spacing is remarkably consistent with previous DNS results. Specifically, \citet{garcia2011hydrodynamic} observed that drag reduction saturated near $s^+ \approx 21$ and performance deteriorated as riblet size increased beyond this critical value. The DNS results of \citet{choi1993direct} also indicate that drag-reducing riblets weaken sweeps and ejections in the near-wall region, thereby reducing turbulent momentum transfer in the wall-normal direction. This observation is consistent with the predicted suppression of the NW resolvent mode.

Of course, there are quantitative differences between the present low-order model predictions and DNS results. For instance, the NW resolvent modes are suppressed by approximately $35\%$ relative to smooth wall values for $s^+ \approx 21$. The maximum drag reduction obtained in DNS by \citet{garcia2011hydrodynamic} was $<10\%$. Further, model predictions show that the gain does not exceed the smooth wall value until riblet size exceeds $s^+ \approx 38$. In contrast, DNS results indicate an increase in drag relative to the smooth wall value for riblets of size $s^+ \approx 30$. One possible reason for this discrepancy at high $s^+$ is the breakdown of the viscous flow regime within the riblet grooves. The model used to compute the mean velocity profile for the resolvent operator is not applicable in this regime. Another potential explanation for the more dramatic deterioration of performance observed in DNS is the emergence of energetic spanwise rollers resembling Kelvin-Helmholtz vortices for large $s^+$. This issue is explored further in the following section.

\begin{figure}[H]
	\centering
	\includegraphics[scale = 0.65]{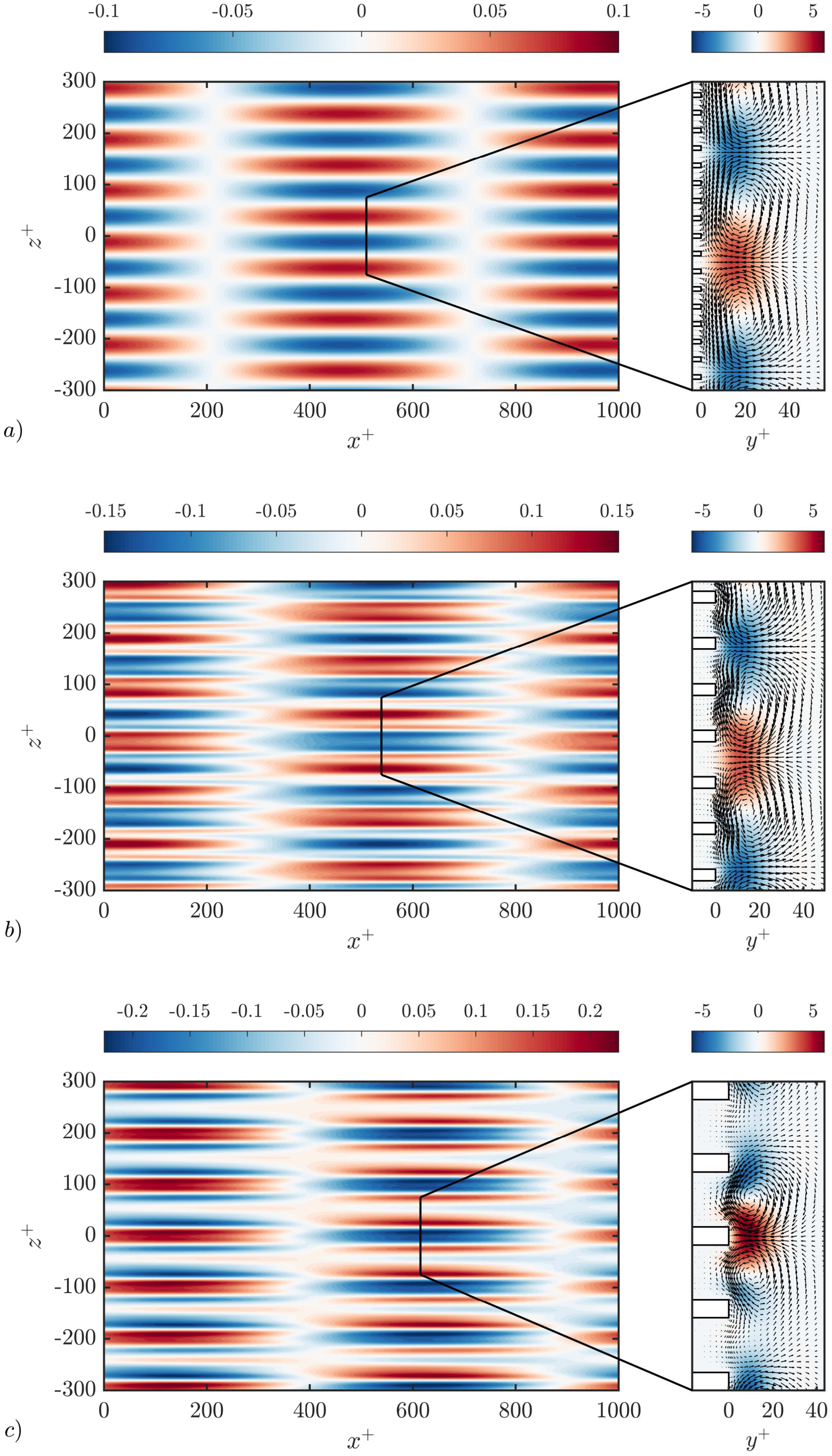}
	\caption{Velocity structure for resolvent modes resembling the NW cycle for riblet spacing: a) $s^+=8$, b) $s^+=21$, and c) $s^+=33$. Figures on the left show the wall normal velocity in the streamwise-spanwise plane at $y^+\approx 6$. Figures on the right depict the flow structure in the spanwise-wall normal plane. The vectors show in-plane velocities ($v,w$) while the red and blue shading shows the out-of-plane streamwise velocity ($u$). 
	}\label{fig:nwc_for_sp}	
\end{figure}

The discussion presented above suggests that the amplification of resolvent modes resembling the NW cycle serves as a useful indicator of performance, even if this measure does not yield quantitative predictions for drag reduction. Next, we evaluate the effect of riblets on the \textit{structure} of resolvent modes resembling the NW cycle. Figure~\ref{fig:nwc_for_sp} shows the predicted mode structure over riblets with spacing $s^+ = 8$, $s^+=21$, and $s^+=33$. The spacings $s^+=8$ and $s^+=33$ represent the limiting cases tested in DNS by \citet{garcia2011hydrodynamic}. Riblet spacing $s^+=21$ corresponds to the optimal size before performance begins to deteriorate. 

Predicted mode structures reveal similar mechanisms to those observed by \citet{suzuki1994turbulent} and \citet{goldstein1998secondary} in numerical simulations and by \citet{lee2001flow} in flow visualization experiments, albeit for triangular or scalloped riblets with sharper tips. Specifically, for $s^+ = 8$ and $s^+ = 21$, the counter-rotating vortices and alternating regions of high and low streamwise velocity associated with the NW resolvent mode are pushed above the riblet tips (fig.~\ref{fig:nwc_for_sp}a,b). There is limited flow penetration into the riblet grooves. Thus, it appears that these riblets effectively limit the wall-normal transfer of momentum by blocking the downwash and crossflows associated with the NW cycle. Recall from fig.~\ref{fig:nwc_for_amp} that this change in flow structure for riblets with $s^+ = 8$ and $s^+ = 21$ is associated with a substantial reduction in amplification for the NW resolvent modes. One potential explanation for this reduction in gain is that riblets limit turbulent energy extraction from the mean flow by pushing the NW cycle away from the wall into a region of lower mean shear. Figures~\ref{fig:nwc_for_sp}a) and b) also show that mode structure remains relatively narrow-banded over the smaller riblets. In other words, most of the energy associated with this resolvent mode is concentrated in the specified wavenumber-frequency combination, $\bkv$. Energy content in the coupled Fourier modes $\bkv \pm n\kss$ is more limited. This concentration of energy is clearly evident in the highly periodic distribution of high and low streamwise velocity observed in the spanwise direction (right-hand side of fig.~\ref{fig:nwc_for_sp}a,b). Physically, this observation is indicative of limited interaction between the NW cycle and the riblets.

In contrast to the flow features discussed above, for riblets with $s^+ = 33$ (fig.~\ref{fig:nwc_for_sp}c) the NW resolvent mode is characterized by increased flow penetration into the riblet groove region. Thus, as size increases, the riblets become less effective at limiting the turbulent transfer of momentum in the wall normal direction, which is consistent with previous observations \citep{choi1993direct,lee2001flow,garcia2011hydrodynamic}. The flow structure is also noticeably less periodic in the spanwise direction. In other words, energy is distributed more evenly across all the different components of the wavenumber-frequency set $\bkv+$. Greater energy transfer into the coupled Fourier harmonics is consistent with previous simulation results which suggest that interaction between large riblets and near-wall turbulence leads to the generation of secondary vorticity near the tips \citep{goldstein1998secondary}. Figure~\ref{fig:nwc_for_amp} shows that for $s^+ = 33$, the riblets are also less effective at reducing gain for the NW resolvent mode. Since the predicted mode structure suggests greater flow penetration into the riblet grooves and increased interaction with additional harmonics of the mean flow (\ref{eqn:mean22_fourier}), this deterioration of performance can be attributed to greater energy extraction from the mean flow.

\subsection{Kelvin-Helmholtz Vortices}\label{sec:kh}

Next, we use the resolvent framework to test for the emergence of spanwise rollers resembling Kelvin-Helmholtz vortices over rectangular riblets. This evaluation is motivated by the DNS results of \citet{garcia2011hydrodynamic}, who observed the emergence of spanwise rollers over rectangular riblets with spacing larger than $s^+ \approx 24$ ($l_g^+ \approx 15$). For these cases, spectra for the Reynolds shear stress and wall-normal velocity showed a strong accumulation of energy for structures with streamwise lengthscale $\lambda_x^+ \approx 150$ and large spanwise extent. A momentum balance showed that the Reynolds stress contribution from structures with $65 \le \lambda_x^+ \le 290$ and $\lambda_z^+ \ge 130$ was responsible for the skin friction increase observed for $s^+ \ge 24$. Further, conditional averaging suggested that these spanwise rollers were centered around $y^+ \approx 10-15$, while cospectra for Reynolds stress showed a peak at $y^+ \approx 5$. The convective speed associated with these structures was estimated to be $c^+ < 8$. To explain the emergence of these spanwise-coherent structures, \citet{garcia2011hydrodynamic} pursued an inviscid, two-dimensional linear stability analysis using both a piecewise-linear velocity profile and an approximation to the turbulent mean profile. Consistent with DNS observations, these analyses showed a sharp transition in stability as riblet size increased above $l_g^+ \approx 11$. However, the streamwise wavelength of the most unstable modes was predicted to be $\lambda_x^+ \approx 80$.

\begin{figure}[H]
	\centering
	\includegraphics[scale = 0.8]{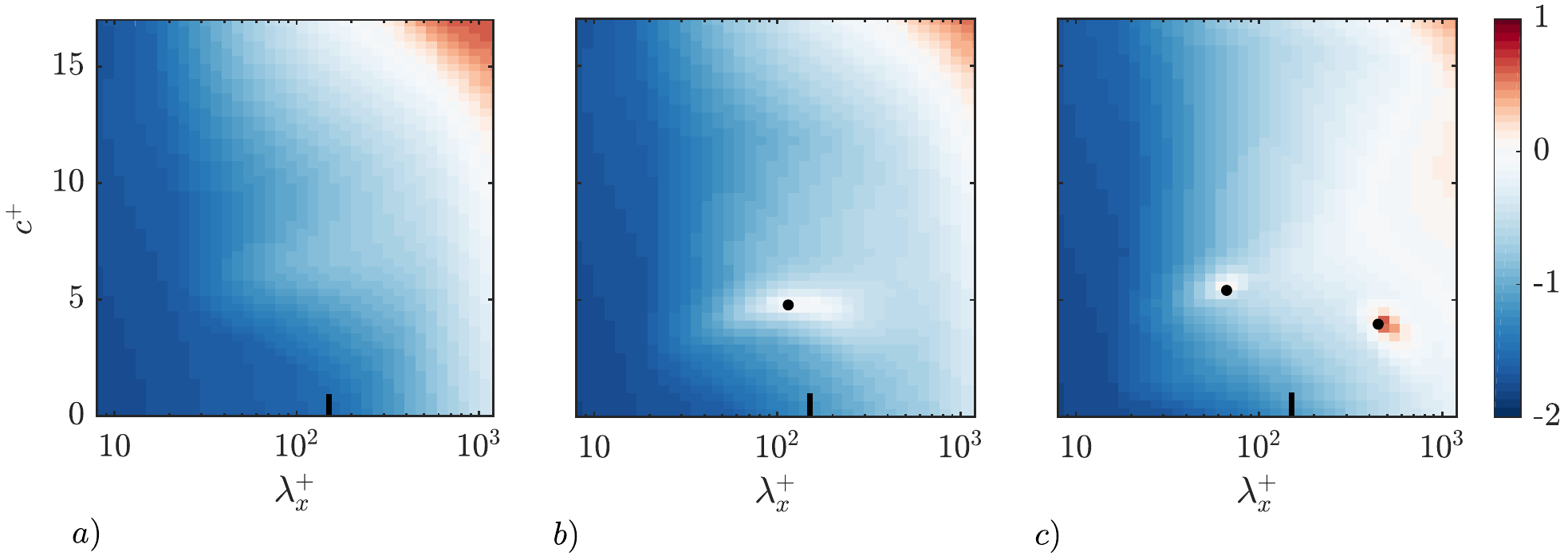}
	\caption{Amplification of spanwise-constant structures as a function of streamwise wavelength and wave speed for riblet spacing: a) $s^+ = 8$, b) $s^+ = 21$, and c) $s^+ = 33$. Red and blue shading represent contours of $\log_{10}(\sigma_{\bkv+})$. The black circles show a local maximum. The vertical bar shows where $\lambda_x^+=150$.} \label{fig:coarse_kh_search}
\end{figure}

Figure~\ref{fig:coarse_kh_search} shows resolvent-based predictions for the gain associated with spanwise-constant structures (i.e., $\kzz = 0$) of varying streamwise wavelength and convective speed. Predictions are shown for riblet spacings $s^+ = 8$, $s^+ = 21$, and $s^+ = 33$. For riblets with spacing $s^+ = 8$, fig.~\ref{fig:coarse_kh_search}a) does not show an obvious local maximum in amplification. For this drag-reducing case, no spanwise rollers were observed in DNS. A local peak in amplification is first observed for $s^+ = 21$, as shown in fig.~\ref{fig:coarse_kh_search}b). For this case, there is a region of high amplification localized between $\lambda_x^+ \approx 90-200$ and $c^+\approx 4-5$. The gain in this region of spectral space is an order of magnitude larger than that predicted for the smaller riblets with $s^+ = 8$. The most amplified structure has wavelength $\lambda_x^+ \approx 130$ and travels at speed $c^+ \approx 4.6$. As riblet size increases, the region of high amplification spreads in spectral space. As an example, for $s^+=24$ the high-gain region extends from $\lambda_x^+ \approx 80-300$ and $c^+\approx 4-6$ (data not shown here). The most amplified structure for $s^+ = 24$ has streamwise wavelength $\lambda_x^+ \approx 200$ and convection speed $c^+\approx 4.5$. Figure~\ref{fig:coarse_kh_search}c) shows that as riblet spacing increases further to $s^+ = 33$, the single region of high amplification evident in fig.~\ref{fig:coarse_kh_search}b) separates into two distinct high-gain zones. The first region corresponds to structures with streamwise wavelength $\lambda_x^+ \approx 70$, while the second region corresponds to longer structures with $\lambda_x^+\approx 450$.

\begin{figure}[H]
	\centering
	\includegraphics[scale = 1]{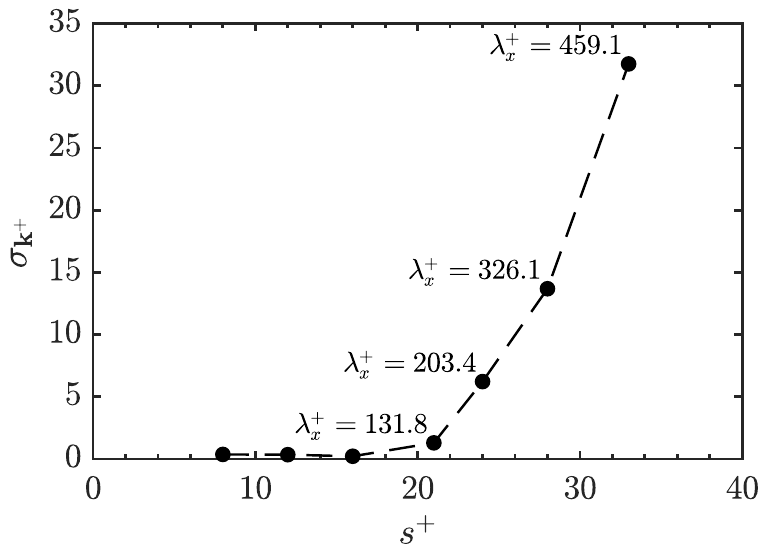}
	\caption{Maximum gain for spanwise-constant resolvent modes resembling Kelvin-Helmholtz vortices as a function of riblet spacing. The inset text shows the streamwise wavelength for the most amplified mode.}\label{fig:kh_for_amp}
\end{figure}

For a direct comparison with the NW resolvent modes considered in the previous section, fig.~\ref{fig:kh_for_amp} shows how the gain associated with the most amplified spanwise-constant structure with streamwise wavelength $50 \le \lambda_x^+ \le 500$ and convection speed $2 \le c^+ \le 10$ varies with riblet size. Consistent with the results shown in fig.~\ref{fig:coarse_kh_search}, a sharp increase in gain is observed as riblet size increases above $s^+ = 21$. This increase in gain also coincides with an increase in the wavelength of the most amplified spanwise roller. A comparison between fig.~\ref{fig:nwc_for_amp} and fig.~\ref{fig:kh_for_amp} shows that the amplification of spanwise-constant structures exceeds that of the NW resolvent mode for $s^+ \ge 24$. For $s^+ = 33$, gain for the spanwise-constant structure ($\sigma_{\bkv+} \approx 33$) is an order of magnitude larger than that for the NW resolvent mode ($\sigma_{\bkv+} \approx 3$).

The predictions shown in fig.~\ref{fig:coarse_kh_search} are in reasonable agreement with DNS results of \citet{garcia2011hydrodynamic}. Although the spanwise rollers first become energetic for $s^+ \approx 24$ in DNS, there is evidence of their appearance at $s^+ = 21$. The wavelength and convection speed of the most amplified resolvent mode for $s^+ = 21$ is slightly lower than that for the structures observed in DNS: $(\lambda_x^+, c^+) \approx (130,4.6)$ vs. $(\lambda_x^+, c^+) \approx (150,6)$. However, the predicted \textit{spreading} of the high-gain region from $\lambda_x^+ \approx 90-200$ for $s^+ = 21$ to $\lambda_x^+ \approx 65-450$ for $s^+ = 33$ is consistent with spectra obtained in DNS. Specifically, premultiplied spectra for wall-normal velocity obtained in DNS show that the high-energy region for structures with large $\lambda_z^+$ widens in $\lambda_x^+$ as riblet size increases. Further, the dramatic increase in gain beyond $s^+\ge 21$ predicted here is consistent with the DNS finding that the spanwise rollers contribute substantially to the observed deterioration of performance.

\begin{figure}[H]
	\centering
	\includegraphics[scale = 0.8]{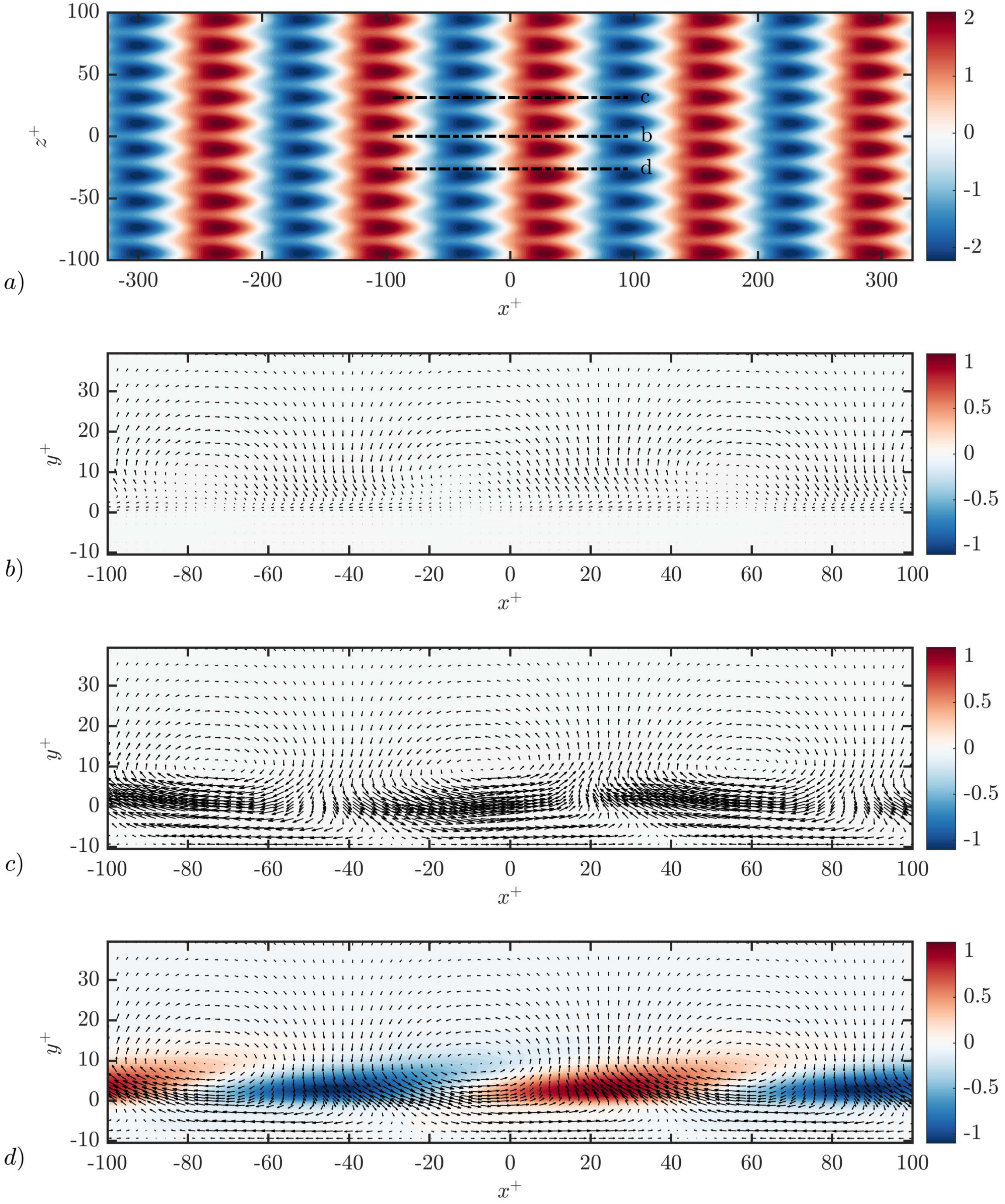}
	\caption{Predicted flow structure for the most amplified spanwise-constant resolvent mode over riblets with spacing $s^+=21$. a) Distribution of wall-normal velocity in a streamwise-spanwise plane at $y^+ \approx 5$. b) Predicted flow field above the riblet tips. c) Predicted flow field in the middle of the riblet grooves. d) Predicted flow field at an intermediate spanwise location. For b)-d) the vectors show the in-plane velocities ($u,v$), while the red and blue shading shows the out-of-plane spanwise velocity ($w$).}\label{fig:kh_instability_sp21}
\end{figure}

Finally, fig.~\ref{fig:kh_instability_sp21} shows the predicted flow structure for the highest-gain resolvent mode at $s^+ = 21$, i.e., the structure with $(\lambda_x^+, c^+) \approx (130,4.6)$ highlighted in fig.~\ref{fig:coarse_kh_search}b). Predictions shown in fig.~\ref{fig:kh_instability_sp21}b)-d) indicate that this resolvent mode resembles a spanwise-constant roller that is centered near $y^+\approx 10$ and does not extend beyond $y^+ \approx 30$. Conditionally-averaged flow fields obtained from DNS by \citet{garcia2011hydrodynamic} exhibited similar characteristics. Further, fig.~\ref{fig:kh_instability_sp21}c) shows that the vertical velocity associated with this resolvent mode penetrates well into the riblet grooves. Penetration into the riblet grooves is also evident in the wall-normal velocity contours shown in fig.~\ref{fig:kh_instability_sp21}a), which exhibit spanwise periodicity that coincides with the riblet spacing. Thus, turbulent structures resembling this high-gain resolvent mode have the potential to generate substantial mixing and momentum transfer in the region occupied by the riblets --- as observed in DNS.

\subsection{Riblet Shape Optimization}\label{sec:riblet_optimization}

\begin{figure}
	\centering
	\includegraphics[scale = 0.8]{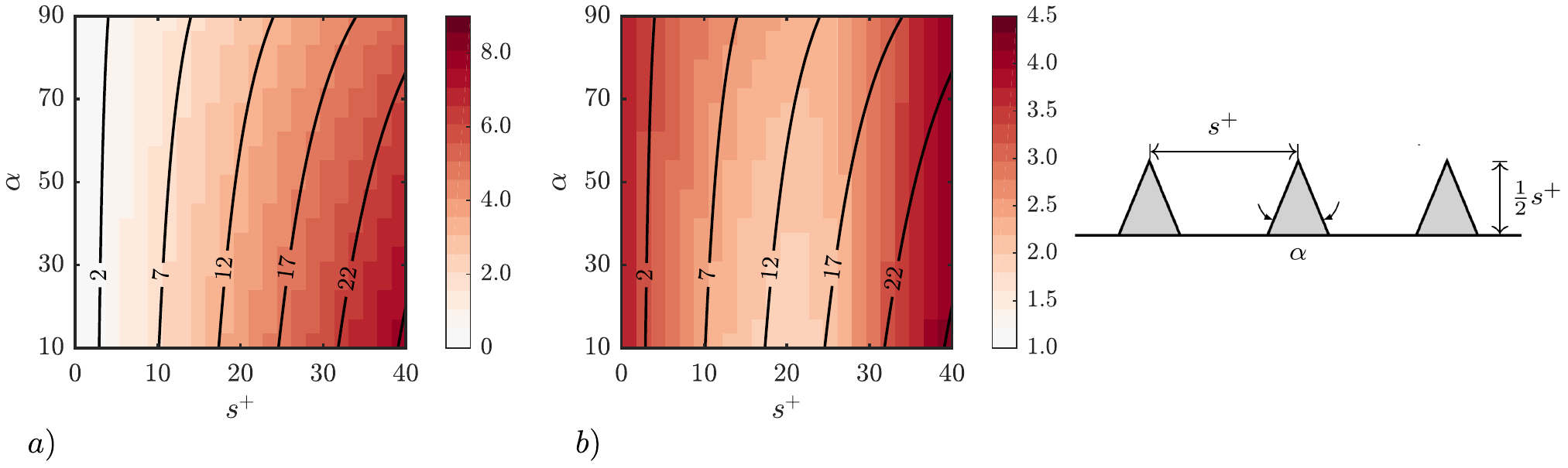}
	\caption{Search for optimal trapezoidal riblet geometries with height to spacing ratio $h/s=0.5$. a) Predictions for the interfacial slip velocity as a function of inner-normalized spacing and ridge angle. b) Predicted gain for resolvent modes resembling the NW cycle. Solid black contours show values for the inner-normalized groove length scale, $l_g^+$. The assumed geometry is shown on the right.}\label{fig:riblet_optimization}
\end{figure}

Model predictions presented in the previous two sections show that the extended resolvent framework developed here is able to reproduce trends observed in DNS with limited computation. In particular, the amplification of a single resolvent mode resembling the energetic NW cycle serves as a useful indicator of drag reduction performance. Of course, this single parameter does not capture all the salient physics dictating riblet performance, such as the emergence of spanwise rollers. However, the emergence of high-gain spanwise-constant structures appears to coincide with minimum gain for the NW mode (see fig.~\ref{fig:nwc_for_amp} and fig.~\ref{fig:kh_for_amp}). Therefore, minimum $\sigma_{\bkv+}$ for the NW mode may serve as a useful metric for the purposes of shape optimization. For proof-of-concept, in this section, we pursue limited shape optimization for triangular, trapezoidal, and rectangular riblets using this metric.

Figure~\ref{fig:riblet_optimization} shows predictions for the interfacial slip velocity (a) and NW mode gain (b) for a family of trapezoidal riblets. The height to spacing ratio for these riblets is fixed at $h/s = 0.5$. The ridge angle and spacing are varied systematically over the following ranges: $\alpha \in [10^\circ,90^\circ]$ and $s^+ \in [0.125,40]$. A total of 576 different configurations are considered in this figure. This extensive exploration is enabled by the relatively low computational cost associated with evaluating a single mode. Importantly, model predictions are broadly consistent with the results compiled by \citet{bechert1997experiments}. For instance, experimental results show that drag reduction increases as the ridge angle decreases, i.e., as the trapezoids approach a blade-like geometry. Further, for fixed $h/s=0.5$, the optimal spacing remained between $s^+ \approx 16-19$ for all the ridge angles considered. The predictions shown in fig.~\ref{fig:riblet_optimization} follow a similar trend. The amplification of NW modes shows a clear minimum around $s^+\approx 20$ for all the different ridge angles considered. In addition, for spacing $s^+ \le 30$, the interfacial slip velocity increases and NW mode gain decreases as the ridge angle, $\alpha$, decreases. An increase in slip velocity and NW mode suppression are both indicative of improved riblet performance. Note that minimum gain for the NW modes over trapezoidal riblets coincides with the contour $l_g^+ \approx 12$ (solid black line in fig.~\ref{fig:riblet_optimization}b).

\begin{figure}[H]
	\centering
	\includegraphics[scale = 0.8]{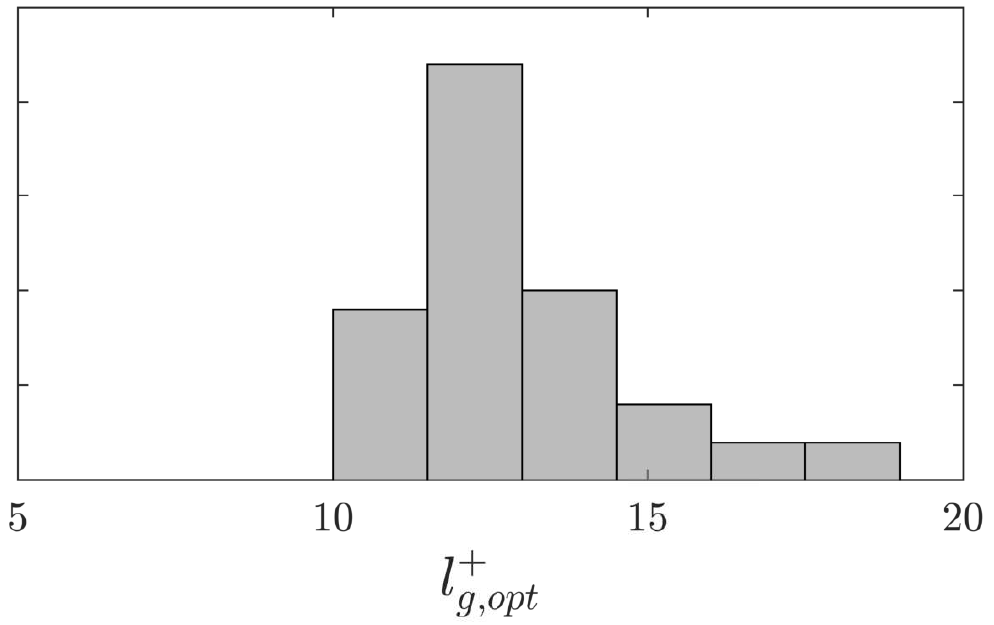}
	\caption{Distribution of the optimal riblet size $l_g^+$, determined from the square root of the groove cross-sectional area for rectangular, triangular, and trapezoidal riblets. The optimum size is identified based on minimum singular values for resolvent modes resembling the NW cycle.}\label{fig:riblet_optimization1}
\end{figure}

We have also pursued similar parametric searches for triangular and rectangular riblets. For brevity, model predictions for all of these cases are not reproduced here. Instead, fig.~\ref{fig:riblet_optimization1} shows a histogram for the optimal $l_g^+$ values identified through this process. These $l_g^+$ values correspond to the spacing that minimizes amplification of the NW resolvent mode for a specific family of shapes (e.g., fixed $\alpha$ for trapezoidal and triangular riblets, fixed $t/s$ ratio for rectangular riblets). The histogram peaks near $l_g^+ \approx 12$, which is unsurprising given the results shown in fig.~\ref{fig:riblet_optimization}b). The mean value of $l_g^+$ for the 54 different optimal geometries represented in this figure is $l_g^+ = 13.0 \pm 1.7$. This mean value is in agreement with the findings of \citet{garcia2011hydrodynamic}, who compiled many previous experimental and numerical datasets to show that optimal drag reduction for a variety of riblet geometries coincides with a value of $l_g^+ \approx 10.7 \pm 1$.

\section{Conclusion}\label{sec:conclusion}
Riblets are arguably some of the simplest --- and most effective --- control solutions proposed thus far for wall-bounded turbulent flows. Simulations, experiments, and theoretical efforts over the past three decades have provided significant insight into the physical mechanisms responsible for drag reduction over riblets. The present paper leverages these physical insight to generate a low-order modeling framework that can guide the design and optimization of riblet geometry. Specifically, the extended resolvent framework developed here allows us to consider the effect of riblets on individual Fourier modes that resemble dynamically-important features of the flow field --- features known to have a direct bearing on riblet performance. 

Model predictions presented in Section~\ref{sec:nw_cycle} showed that the variation in gain of a \textit{single} resolvent mode resembling the NW cycle over riblets reproduced drag reduction trends from DNS. Moreover, the predicted changes in flow structure for this single mode were consistent with prior observations in simulations and experiments \citep{choi1993direct,goldstein1998secondary,lee2001flow}. The low computational cost associated with this single mode evaluation enabled the preliminary shape optimization effort presented in Section~\ref{sec:riblet_optimization}. Recent work by \citet{garcia2011hydrodynamic} shows that the emergence of spanwise rollers resembling Kelvin-Helmholtz vortices is another pathway that leads to riblet performance deterioration. Section~\ref{sec:kh} confirmed that the resolvent framework is also able to predict the emergence of such flow structures. 

Of course, the modeling framework developed here does have limitations. One key limitation is the requirement of a mean velocity profile. The model employed here was able to reproduce prior mean profile observations reasonably well. However, this may not be the case for all flow conditions and riblet geometries. Related to this point, evaluation of model prediction sensitivity to the mean profile is also needed. Further, the volume penalization framework employed here does not reproduce the riblet geometry exactly. In particular, the truncated Fourier series representation shown in (\ref{eqn:wall_model}) does not yield rapid convergence for the sharp volume penalization transitions needed to accurately represent the riblets. This limitation could be remedied by extending the resolvent framework to more sophisticated immersed boundary approaches. Also, we only considered relatively simple narrow-banded surface geometries in this paper, i.e., spanwise-periodic and streamwise constant riblets. Extension to account for multi-scale or 3D surface features would require inclusion of a greater number of coupled Fourier harmonics in the extended resolvent operator (\ref{eqn:modified_resolvent_f}), increasing computational expense. Despite these limitations, the model verification and preliminary optimization effort described in this paper shows that the extended resolvent framework can serve as a useful low-order design tool prior to testing in more expensive simulations or experiments. \\

This material is based on work supported by the Air Force Office of Scientific Research under award FA9550-17-1-0142 (program manager Gregg Abate).
\newpage
\bibliographystyle{plainnat}
\bibliography{chavarin_luhar_AIAA_refs}
\end{document}